\documentclass[12pt]{article}

\begin{document}

\title{
The Finite Size $SU(3)$ Perk-Schultz Model 
 with  Deformation Parameter $q=\exp(\frac{2 i \pi}{3})$}
\author{F.~C.~Alcaraz$^{\rm a}$, Yu.~G.~Stroganov$^{\rm b,c}$\\
\small \it $^{\rm a}
$Universidade de S\~ao Paulo, 
Instituto de F\'{\i}sica de S\~ao Carlos, \\
\small \it C.P. 369,13560-590, S\~ao Carlos, SP, Brazil \\
\small \it $^{\rm b}$Departamento de F\'{\i}sica, Universidade 
Federal de S\~ao Carlos, 13565-905, \\
\small \it S\~ao Carlos, SP, Brazil \\
\small \it $^{\rm c}$Institute for High Energy Physics\\[-.5em]
\small \it 142284 Protvino, Moscow region, Russia}
\date{}

\maketitle

\begin{abstract}
From extensive  numeric diagonalizations of the $SU(3)$  Perk-Schultz 
Hamiltonian with a special
value of the anisotropy and different boun-\\dary conditions, 
we have observed simple regularities
for a significant part of its eigenspectrum. 
In particular the ground state energy and nearby excitations
 belong to this part of the spectrum.
 Our simple formulae describing these regularities remind, 
apart from some selection rules,  
the eigenspectrum  of the free fermion
model. 
Based on the numerical observations  we formulate several conjectures. 
Using explicit solutions of the associated nested Bethe-ansatz equations, 
guessed from an analysis of the functional equations of the model, 
we provide evidence for a part of them.
 
\end{abstract}

\vskip 1em

\begin{center}{\bf1. Introduction}
\end{center}

Since the pioneering work of H.  Bethe in 1931 the Bethe ansatz 
and its generalizations have proved to be  quite efficient tools 
in the description
of the eigenvectors of a huge variety of 
one-dimensional quantum chains 
 and two-dimensional transfer matrices (see, e. g., \cite{rev} 
for reviews). 
Models with wave function given by this ansatz are 
considered exactly integrable. 
According to the Bethe ansatz the amplitudes of the wave 
functions are expressed in terms of a sum of plane waves 
whose wave numbers are given in terms of non-linear and highly non 
trivial coupled equations known as the Bethe-ansatz equations (BAE). 
These equations in several cases, thanks some appropriate guessing 
on the topology of roots, are solvable in the thermodynamic limit 
providing the understanding of the large-distance physics.

However the exact integrability is a property independent of the 
lattice size and the exact solution of the associated BAE for 
finite chains is an important step toward the complete mathematical
 and physical understanding of these models. Due to the 
complexity of the BAE up to our knowledge, only in two special 
cases some of the solutions are known analytically, namely, 
the trivial 
free-fermion case and the XXZ chain at the special anisotropy 
$\Delta=-1/2$~\cite{alc1,YGS}.
The solution in this last case is obtained by exploring the 
functional relations of the model~\cite{YGS}. Even in the last case, 
although several exact properties of the wave function were 
conjectured~\cite{ALL} a complete and closed calculation of their 
amplitudes is still missing.  
In this paper we  are going to present a new set of analytical 
solutions of BAE for finite chains. These solutions correspond to 
BAE of the anisotropic $SU(3)$ Perk-Schultz model~\cite{PS}, 
or the  anisotropic $SU(3)$ Sutherland  model~\cite{Sutherland}, at a special 
value of the anisotropy. Contrary to the XXZ case the Bethe ansatz 
for this model is of nested type  and the solution are going to be 
derived by generalizing the functional method originally applied 
to the XXZ chain. 

The paper is organized as follows.
In section 2 we give the main definitions and formulate the 
corresponding BAE.
In section 3 we state a set of conjectures that were obtained from 
extensive "experimental" work on exact bruteforce diagonalizations 
of the quantum chains.
In section 4 we derive, for the Hamiltonian with periodic boundary 
conditions the functional relations and at a special value of the 
anisotropy some solutions for the eigenspectra are derived. In section 
5 we present and test directly a set of solutions of the BAE, and 
explain partially the conjectures announced in section 3. Finally 
in section 6 we present our conclusions and a summary of our results.

\vskip 1em

\begin{center}{\bf2. The $SU(3)$ Perk-Schultz model}
\end{center}

The $SU(3)$ Perk-Schultz model~\cite{PS} is the anisotropic 
version of the 
$SU(3)$ Sutherland model~\cite{Sutherland} with Hamiltonian, 
in a L-sites chain, given by 

\begin{eqnarray}
\label{H}
&&H_{q}^{p} =\sum_{j=1}^{L-1} H_{j,j+1} + p H_{L,1}\qquad (p=0,1), 
\qquad \mbox{where}  \\ 
&&H_{i,j} =-\sum_{a=0}^{1} \sum_{b=a+1}^{2} 
( E_i^{ab} E_{j}^{ba} + E_i^{ba} E_{j}^{ab}- q E_i^{aa} E_{j}^{bb} 
- 1/q  E_i^{bb} E_j^{aa} ) \nonumber
\end{eqnarray}
The $3 \times 3$ matrices $E^{ab}$ have elements $(E^{ab})_{cd}=
\delta^a_c \delta^b_d$ and $q=\exp (i \eta)$ is the anisotropy of 
the model.
The cases of free and periodic boundary conditions are obtained 
by setting $p=0$ and $p=1$ in (\ref{H}), respectively. 
This Hamiltonian describe the dynamics of a system containing 
three classes of particles (0,1,2)
 with on-site hard-core exclusion. At $q=1$ the model is 
$SU(3)$ symmetric and for $q \ne 1$ the model has a $U(1) 
\bigotimes U(1)$ symmetry due to the conservation of the 
number of particles of each specie. Consequently we can 
separate the Hilbert space into block disjoint sectors 
labeled by $(n_0,n_1,n_2)$, where $n_i=0,1,...,L$ is the number 
of particle of specie i (i=0,1,2). The Hamiltonian has a $S_3$ 
symmetry due to its invariance under the permutation of distinct  
species, that implies that all the energies can be obtained from the 
sectors $(n_0,n_1,n_2) $, where $n_0 \le n_1 \le n_2$ and  
$n_0+n_1+n_2=L$. Moreover in the special case of free boundaries 
($p=0$), the quantum chain (\ref{H}) is also invariant under the 
additional quantum $SU(3)_q$ symmetry implying that all energies 
in the sector $(n_0^{\prime},n_1^{\prime},n_2^{\prime})$ with 
$n_0^{\prime} \le n_1^{\prime} \le n_2^{\prime}$  are degenerated 
with the energies belonging to the sectors 
$(n_0,n_1,n_2)$ with $n_0 \le n_1 \le n_2$, 
if $n_0^{\prime} \le n_0$ and
$n_0^{\prime} + n_1^{\prime} \le n_0 +n_1$.

The eigenenergies of the Hamiltonian (\ref{H}) for 
$p=0$ or $p=1$ in the sector $(n_0,n_1,n_2)$ are given by 
\begin{equation}
\label{ba1}
E=-\sum_{j=1}^{n_0+n_1} \biggl(-q-\frac{1}{q}+\frac{\sin(u_j-\eta /2)}
{\sin(u_j+\eta /2)} +\frac{\sin(u_j+\eta /2)}
{\sin(u_j-\eta /2)} \biggr),
\end{equation}
where the variables $\{u_j, j=1,2,...,n_0+n_1\}$ and 
the auxiliary variables 
$\{v_j, j=1,2,...,n_0\}$ are the roots of the coupled Bethe ansatz. 
These BAE are of nested type and in the case of periodic boundary 
they are given by (see e. g. \cite{resh,devega})
\begin{eqnarray}
\label{ba2}
&&\biggl[\frac{\sin(u_k+\eta /2)}{\sin(u_k-\eta /2)}\biggr]^L=
-\prod_{i=1}^{n_0+n_1} \frac{\sin(u_k-u_i+\eta )}{\sin(u_k-u_i-\eta )} 
\> \prod_{j=1}^{n_0} \frac{\sin(u_k-v_j-\eta /2)}{\sin(u_k-v_j+\eta /2)}, 
\nonumber \\
&&\prod_{i=1}^{n_0} \frac{\sin(v_l-v_i+\eta )}{\sin(v_l-v_i-\eta )} \> 
\prod_{j=1}^{n_0+n_1} \frac{\sin(v_l-u_j-\eta /2)}{\sin(v_l-u_j+\eta /2)}=-1,
\end{eqnarray}
where $k=1,...,n_0+n_1$ and $l=1,...,n_0$.

In the case of free boundary the BAE are given by~\cite{freeBAE}
\begin{eqnarray}
\label{ba3}
&&\biggl[\frac{\sin(u_k+\eta /2)}{\sin(u_k-\eta /2)}\biggr]^{2L} \> 
\prod_{i=1}^{n_0} \frac{\sin(u_k+v_i+\eta /2) \sin(u_k-v_i+\eta /2)} 
{\sin(u_k+v_i-\eta /2) \sin(u_k-v_i-\eta /2)}=\nonumber \\
&&\prod_{j=1, j \ne k}^{n_0+n_1} \frac{\sin(u_k+u_j+\eta ) 
\sin(u_k-u_j+\eta )} {\sin(u_k+u_j-\eta ) \sin(u_k-u_j-\eta )} \> , \nonumber \\
&&\prod_{i=1,i \ne l}^{n_0} \frac{\sin(v_l+v_i+\eta ) 
\sin(v_l-v_i+\eta )} {\sin(v_l+v_i-\eta ) \sin(v_l-v_i-\eta )} = \nonumber \\
&& \prod_{j=1}^{n_0+n_1} \frac{\sin(v_l+u_j+\eta /2) 
\sin(v_l-u_j+\eta /2)} {\sin(v_l+u_j-\eta /2) \sin(v_l-u_j-\eta /2)},
\end{eqnarray}
where $k=1,...,n_0+n_1$ and $l=1,...,n_0$.
In the case of periodic boundaries the momentum $P=\frac{2 \pi l}{L}$ 
$(l=0,1,...,L-1)$ of the associated eigenstate is given by 
\begin{equation}
\label{ba4}
\exp(i P) =\prod_{k=1}^{n_0+n_1} \frac{\sin(u_k-\eta /2)}{\sin(u_k+\eta /2)}.
\end{equation}

 The solutions of the BAE are going to provide the eigenenergies of 
(\ref{H}) if they correspond to non-zero norm Bethe states. 
The combinatory nature of the Bethe wave functions imply that solutions 
of (\ref{ba2}) or (\ref{ba3}) with coinciding roots 
produce null states. 
Nevertheless the requirement of non-coinciding roots does not necessarily 
ensure a  genuine eigenvector, since conspicuous cancellation even in this 
case, can render a vector with null norm.
Although all eigenenergies of the Hamiltonian can be obtained, apart from 
predicted degeneracies, by restricting to the sectors $(n_0,n_1,n_2)$ with   
$n_0 \le n_1 \le n_2$, the Bethe ansatz implementation in its coordinate 
version is valid for arbitrary values of $n_0,n_1,n_2$. 
However as we shall see in section 5, several solutions with non-coinciding 
roots for sectors out of the range $n_0 \le n_1 \le n_2$, but corresponding
 to null state can be obtained. In fact even in the XXZ chain, 
where the BAE are simpler, solutions with non-coinciding roots 
\footnote{We have found, for example, a continuous set of non-coinciding roots 
of the BAE for the periodic XXZ with $L=4$ sites and $n=3$. All this set of 
solutions give us zero eigenvectors (see also~\cite{PrStr} for further 
considerations).}
that correspond to null-norm states can be obtained when the number of 
roots $n$ is out of the range $n \le L/2$. 
In the case of the BAE for the XXZ chain, in a recent 
paper~\cite{Baxter} Baxter gives strong evidence that with 
suitable parametrization the entire eigenspectra can be obtained 
from the non-coinciding roots of the associated BAE in the sector 
with the number of roots $n \le L/2$.
In a similar way we are going to assume on this paper that all distinct 
eigenenergies of (\ref{H}) can be obtained from the solutions 
$\{u_i\},\{v_i\}$, with non-coinciding values in the sectors 
when $n_0 \le n_1 \le n_2$.
Conversely solutions in sectors out of this range, should necessarily 
be degenerated with energies occurring in sectors within the range, 
if they do not correspond to zero-norm states.

\vskip 1em
\begin{center}{\bf3. Conjectures merged from numerical studies}
\end{center}

In this section we state a series of conjectures that are consistent 
with the exact brute-force diagonalization of the Hamiltonian 
(\ref{H}) with free ($p=0$)
and periodic ($p=1$) boundary condition at $q=\exp(2 i \pi/3)$. 
Some of these conjectures are going to be proved in the following sections. 
Let us consider separately the case of periodic and free boundaries.

\vskip 1em

\noindent {\bf 3a - Periodic chain.}

\noindent CONJECTURE 1.  The Hamiltonian (\ref{H}) with L sites at $q=\exp(2 i \pi/3)$
has eigenvectors (not all of them) with energy and momentum given by 
\begin{eqnarray}
\label{C1}
&&E_I=-\sum_{j \in I} (1+2 \cos \frac{2 \pi j}{L}),  \\
\label{C2}
&& P_I=\frac{2 \pi}{L} \sum_{j \in I} j,
\end{eqnarray}
with $I$ being a subset of $\cal I $ unequal elements of the 
set  $\{1,2,...,L\}$.
The number $\cal I$ has to be odd ${\cal I} =2k+1$ and the sector of appearance of the above levels  is  $S_k \equiv (k,k+1,L-2k-1)$, $0 \le k \le (L-1)/2$.

The lowest eigenenergy among the above conjectured values (\ref{C1})
is obtained for the particular set 
$I_0^{(k)}=\{1,2,...,k\} \cup \{L-k,...,L\}$,  
since in this case 
  all contributions $- (1+2 \cos \frac{2 \pi j}{L})$ to (\ref{C1}) 
has the lowest possible values. The corresponding  eigenstate has zero momentum and energy given by
\begin{equation}
\label{C3}
E_0^{(k)}=-\sum_{j \in I_0^{(k)}} (1+2 \cos \frac{2 \pi j}{L}) = -2k-1-2
 \frac{\sin(\pi (2k+1)/L)}{\sin(\pi /L)}.
\end{equation}
 
\noindent CONJECTURE 2. For arbitrary $L=3n+l$ $(l=1,2,3)$, 
the eigenenergy $E_0^{(n)}$
is the lowest one in the sector $S_n$, moreover if $l \ne 3, (L \ne 3n)$ 
it is the ground-state energy of the model.

\vskip 1em

\noindent {\bf 3b - Free boundaries.}

In order to state our conjectures let us define again the special set of 
sectors  of the Hamiltonian (\ref{H}) with $p=0$:

\begin{equation}
\label{C4}
S_k=(\mbox{Int}(\frac{k}{2}), \mbox{Int}(\frac{k+1}{2}), L-k),
\qquad k=0,1,...L.
\end{equation}
 Due to the quantum symmetry $SU(3)_q$, distinct sectors show the 
same eigenenergies.
  For example, for $L=7$ the sectors are
\begin{eqnarray}
\label{Sectors}
&S_0=(0,0,7) &S_1=(0,1,6) \nonumber \\
&S_2=(1,1,5) &S_3=(1,2,4) \nonumber \\
&S_4=(2,2,3) &S_5=(2,3,2) \nonumber \\
&S_6=(3,3,1) &S_7=(3,4,0) \nonumber 
\end{eqnarray}
and we have a special ordering
\begin{equation}
\label{C5}
S_0 \subset S_1 \subset S_2 \subset S_3 \subset S_4 
\equiv S_5 \supset S_6 \supset S_7 .
\end{equation}
This means that, for example, all eigenvalues in sector $S_2$ can also 
be found in sectors $S_3$, $S_4$ and $S_5$, and on the other side all 
eigenvalues in sector $S_7$ also appear in  sectors $S_6$, $S_5$ and $S_4$. 
Sectors $S_4$ and $S_5$ are totally equivalent. 
Let us call in this example the sectors  $S_0,S_1,S_2,S_3,S_4$ as the 
LEFT sectors and $S_5,S_6,S_7$ as the RIGHT ones.
This can be directly generalized to any $L=3n+1$ or $L=3n+2$, 
obtaining $L-n$ left sectors and $n+1$ right ones. 
In the case where $L=3n$ the sectors $S_k$ with $k=0,1,...,2n-1$ and 
$k=2n+1,...,L$ as the left and right sectors, respectively. 
The sector $S_{2n}=(n,n,n)$ is degenerated with two sectors 
$S_{2n-1}=(n-1,n,n+1)$ and $S_{2n+1}=(n,n+1,n-1)$  
$( S_{2n-1} \equiv S_{2n+1})$
and can be considered either as a left or right sector. 
We state now the conjecture.

\noindent CONJECTURE 3. Let $L=3n+l\> (l=0,1,2)$. 
Then the Hamiltonian (\ref{H}) with free boundaries 
$(p=0$ at $q=\exp(2 i \pi/3)$ has eigenvectors with energies given by 
\begin{equation}
\label{C6}
E_I=-\sum_{j \in I} (1+2 \cos \frac{\pi j}{L}),  
\end{equation}
with $I$ is an arbitrary subset formed by k distinct elements of 
the set $\{1,2,...,L-1\}$. Moreover  if $S_k$ is a left sector, 
then these eigenvalues appear in the sectors 
$S_k,S_{k+1},...,S_{L-n}$ $(S_{L-n+1}\> \mbox{for}\> l=0)$, 
and if $S_k$ is a right sector, the eigenvalues appear 
in the sectors $S_{L-n-1},S_{L-n},...,S_{k+1}$.

As a consequence of conjecture 3 the 
Hamiltonian (\ref{H}) has the special eigenvalues
\begin{equation}
\label{C7}
E^{(k)}=-\sum_{j=1}^k (1+2 \cos \frac{\pi j}{L}) = 
1-k- \frac{\sin(\pi (2k+1)/2L)}{\sin(\pi /2L)}
\end{equation}
  and we are now in condition to formulate a remarkable conjecture.

\noindent CONJECTURE 4.  
The lowest energy in the sector $S_k$ is $E^{(k)}$ or $E^{(k-1)}$
 depending if $S_k$ is a left or a right sector respectively.

The minimal value of $E^{(k)}$ is obtained for $k=L-n-1$ and 
our "numerical experiments" induce the conjecture:

\noindent CONJECTURE 5.  The ground-state energy of the Hamiltonian 
(\ref {H}) with free boundary at $q=\exp(2 i \pi /3)$ is given by
\begin{equation}
\label{C8}
E_0=E^{(L-n-1)}= 2-L+n- \frac{\sin(\pi (2n+1)/2L)}{\sin(\pi /2L)}.
\end{equation}

\vskip 1em

\begin{center}{\bf4. Functional relations for the anisotropic $SU(3)$ 
Perk-Schultz model}
\end{center}

 We are going to obtain analytically some of the 
conjectured results presented in the previous section.
Let us consider initially the periodic case when $p=1$ in 
the Hamiltonian (\ref{H}). The eigenenergies 
in the sectors with "particle numbers" $(n_0,n_1,n_2)$ are 
given by (\ref{ba1}) where the Bethe roots   
 $\{u_j, j=1,2,...,n_0+n_1 \equiv m_2\}$ and 
$\{v_j, j=1,2,...,n_0\equiv m_1\}$ are obtained by solving the BAE
 (\ref{ba2}).
Bellow, to simplify the notation,  
we write $\lambda_j^{(1)}$ and $\lambda_j^{(2)}$ 
instead of $v_j$ and $u_j$, respectively.

Defining the pair of sine-polynomials
\begin{equation}
\label{f1}
Q_l(\lambda) = \prod_{j = 1}^{m_l} \sin (\lambda -
\lambda_j^{(l)}),\quad l=1,2,
\end{equation}
the BAE (\ref{ba2}) can be rewritten as
\begin{eqnarray}
&&Q_1(\lambda_j^{(1)}+\eta)\> Q_2(\lambda_j^{(1)}-\eta/2)+ 
Q_1(\lambda_j^{(1)}-\eta) \>Q_2(\lambda_j^{(1)}+\eta/2) = 0 \nonumber \\
\label{f2}
&&\quad (j = 1,2, \ldots m_1),  \\
&& \sin ^L(\lambda_k^{(2)}+\eta/2)\> Q_1(\lambda_k^{(2)}+\eta/2)\>
Q_2(\lambda_k^{(2)}-\eta) + \nonumber \\
&& +\sin  ^L(\lambda_k^{(2)}-\eta/2)\> Q_1(\lambda_k^{(2)}-\eta/2)\>
Q_2(\lambda_k^{(2)}+\eta) = 0 \nonumber \\
\label{f3}
&&\quad  (k = 1,2, \ldots m_2).
\end{eqnarray}
Since from the definitions (\ref{f1}) $Q_l(\lambda_j^{(l)})=0$
for any Bethe roots $\lambda_j^{(l)} (l=1,2)$, 
we should have the functional relations
\begin{eqnarray}
\label{f4}
&&Q_1(\lambda+\eta)\> Q_2(\lambda-\eta/2)+Q_1(\lambda-\eta)\>
Q_2(\lambda+\eta/2) = T_2(\lambda) Q_1(\lambda),  \\
&&\sin ^L(\lambda+\eta/2)\> Q_1(\lambda+\eta/2)\> 
Q_2(\lambda-\eta)+ \nonumber \\
\label{f5}
&&+ \sin ^L(\lambda-\eta/2)\> Q_1(\lambda-\eta/2)\> 
Q_2(\lambda+\eta) = T_1(\lambda) Q_2(\lambda), 
\end{eqnarray}
where $T_2(\lambda)$ and $T_1(\lambda)$ are unknown 
sine-polynomials of the order $m_2$ and $L+m_1$, respectively. 
Shifting $\lambda \rightarrow \lambda \mp \eta/2$ in (\ref{f4}) and inserting the result in (\ref{f5}) we obtain
\begin{eqnarray}
\label{f6}
&& \sin ^L(\lambda \mp \eta/2)\> Q_2(\lambda \pm \eta) + \sin ^L (\lambda \pm \eta/2)\> T_2(\lambda \mp \eta/2)\} Q_1(\lambda \mp
\eta/2)=\nonumber\\
&&\sin ^L (\lambda \pm \eta/2)\> Q_1(\lambda \mp
3\eta/2)+T_1(\lambda)\} Q_2(\lambda).
\end{eqnarray}
We now suppose that $Q_1(\lambda \pm \eta/2)$ and  $Q_2(\lambda)$ have no common roots, in this case:
\begin{eqnarray}
\label{f7}
&&\sin ^L (\lambda \mp \eta/2) Q_2(\lambda \pm \eta)
+ \sin ^L (\lambda \pm \eta/2) T_2(\lambda \mp \eta/2)\} = \nonumber \\
&& = T^{\pm}(\lambda)
Q_2(\lambda),\\
\label{f8}
&&\sin ^L (\lambda \pm \eta/2)\> Q_1(\lambda \mp
3\eta/2)+T_1(\lambda) = T^{\pm}(\lambda) Q_1(\lambda \mp \eta/2), 
\end{eqnarray}
where  $T^{\pm}(\lambda)$ are sine-polynomials\footnote{
These polynomials are the eigenvalues of the transfer 
matrices corresponding to
the  fundamental representations 
of $SU(3)$ in the auxiliary space.} of the degree $L$.
The subtraction of equations (\ref{f8}) among themselves give us  
\begin{eqnarray}
\label{f9} 
&&\sin ^L (\lambda+\eta/2)\>Q_1(\lambda-3\eta/2)-T^{+}(\lambda)
Q_1(\lambda-\eta/2)\nonumber\\
&&+T^{-}(\lambda) Q_1(\lambda+\eta/2)-\sin ^L 
(\lambda-\eta/2)\>Q_1(\lambda+3\eta/2)=0. 
\end{eqnarray}
Similarly both equations (\ref{f7}) give us the relation
\begin{eqnarray}
\label{f10}
&&\sin ^L (\lambda) \sin ^L 
(\lambda+\eta)\>Q_2(\lambda-3\eta/2)\nonumber \\
&&-\sin ^L (\lambda+\eta)\> T^{-}(\lambda-\eta/2)\> 
Q_2(\lambda-\eta/2)\nonumber \\
&&+\sin ^L (\lambda-\eta)\> 
T^{+}(\lambda+\eta/2)\> Q_2(\lambda+\eta/2)\nonumber \\
&& -\sin ^L (\lambda) \sin ^L (\lambda-\eta)\> Q_2(\lambda+3\eta/2)=0.
\end{eqnarray}

Up to now our relations are valid for arbitrary values of the anisotropy $\eta$ and we now are going to restrict to the particular case $\eta=2 \pi /3$ ($q=\exp(2 i \pi/3$), where the several conjectures announced in Section 3 were expected to be valid. At this special value of the anisotropy we have the symmetry
\begin{equation}
\label{f11}
Q_l(\lambda-3\eta/2) = Q_l(\lambda-\pi) = Q_l(\lambda+\pi) = Q_l(\lambda+
3\eta/2)   \quad l=1,2,
\end{equation}
and equations (\ref{f9}) and (\ref{f10}) are given by
\begin{equation}
\label{f12}
\phi (\lambda)\> Q_1(\lambda-\pi) - T^{+} (\lambda)\> Q_1(\lambda-\pi/3) 
+ T^{-} (\lambda) \>Q_1(\lambda+\pi/3)=0,
\end{equation}
and
\begin{eqnarray}
\label{f13}
&&-\sin ^L (\lambda) \phi (\lambda - \pi)\> 
Q_2(\lambda-\pi) - \sin ^L (\lambda+2\pi /3) T^{-} 
(\lambda-\pi /3)\> Q_2(\lambda-\pi/3) \nonumber \\
&& + \sin ^L (\lambda - 2\pi /3) T^{+} (\lambda +\pi /3) \>Q_2(\lambda+\pi/3)=0,
\end{eqnarray}
where
\begin{equation}
\label{f14}
\phi (\lambda)\>= \sin ^L (\lambda+\pi /3) - \sin ^L (\lambda-\pi /3). 
\end{equation}

The shifting $\lambda \rightarrow \lambda \pm 2\pi/3$ in (\ref{f12}) and (\ref{f13}) show that these equations are equivalent to the linear matricial equations
\begin{equation}
\label{f15}
\left|\begin{array}{ccc}
 \phi(\lambda)  & -T^{+}(\lambda) & T^{-}(\lambda) \\
 T^{-}(\lambda +\frac{2\pi}{3})& \phi (\lambda +\frac{2\pi}{3})&
- T^{+} (\lambda +\frac{2\pi}{3}) \\
 -T^{+}(\lambda -\frac{2\pi}{3})&  T^{-} (\lambda -\frac{2\pi}{3})&
\phi (\lambda -\frac{2\pi}{3}) 
\end{array} \right|
\left|\begin{array}{c}
Q_1(\lambda - \pi) \\
Q_1(\lambda - \frac{\pi}{3}) \\ 
Q_1(\lambda + \frac{\pi}{3})
\end{array} \right| = 0,
\end{equation}
and
\begin{equation}
\label{f16}
\left|\begin{array}{ccc}
 \phi(\lambda - \pi)  &  T^{-}(\lambda-\frac{\pi}{3}) & -T^{+}(\lambda + \frac{\pi}{3}) \\
 - T^{+}(\lambda - \pi) & -\phi (\lambda -\frac{\pi}{3})&
T^{-} (\lambda +\frac{\pi}{3}) \\
 T^{-}(\lambda -\pi)& -T^{+} (\lambda -\frac{\pi}{3})&
\phi (\lambda +\frac{\pi}{3}) 
\end{array} \right|
\left|\begin{array}{c}
{\tilde Q}_2 (\lambda - \pi) \\
{\tilde Q}_2(\lambda - \frac{\pi}{3}) \\ 
{\tilde Q}_2(\lambda + \frac{\pi}{3})
\end{array} \right| = 0.
\end{equation}
respectively. 
 In (\ref{f16}) we defined the  new function ${\tilde Q}_2(\lambda) 
= \sin^L (\lambda)\> Q_2(\lambda)$.
It is clear  that $T_2(\lambda+\pi) = T_2(\lambda-\pi)$ and 
consequently from (\ref{f7}) $T_{\pm}(\lambda+\pi) = 
T_{\pm}(\lambda-\pi)$. Equations (\ref{f15}) and 
(\ref{f16}) imply that non trivial solutions are obtained if 
the determinants of the matrices appearing in those equations 
vanish. Actually, by shifting $\lambda \rightarrow \lambda + \pi$ 
in the determinant coming from  (\ref{f16}) we clearly see that this
 last determinant
 vanishes if the one coming from 
(\ref{f15}) also vanishes. 

The calculation of the general solutions $T^{\pm}(\lambda)$ 
that render a null determinant is a quite difficult task, 
however  simple solutions can be obtained (rank 1) by 
imposing a proportionality between the columns of the matrix 
generating the 
determinant\footnote{The  idea to consider decreased rank in the 
functional relations was used previously in \cite{rank1} to 
explain simple energy levels of a special case of the XXZ chain.}, i.e., 
\begin{eqnarray}
\label{f17}
&&\frac{\phi (\lambda)}{-T^+(\lambda)}=
\frac{T^-(\lambda+2\pi /3)}{\phi(\lambda+2\pi /3)}=
\frac{-T^+(\lambda-2\pi /3)}{T^-(\lambda-2\pi /3)}, \nonumber \\
&&\frac{-T^+(\lambda)}{T^-(\lambda)}=\frac{\phi(\lambda+2\pi /3)}
{-T^+(\lambda+2\pi /3)}=\frac{-T^-(\lambda-2\pi /3)}
{\phi(\lambda-2\pi /3)}.
\end{eqnarray}

We can verify the above relations are equivalent to the 
independent equations
\begin{equation}
\label{f18}
T^+(\lambda)\>T^-(\lambda+2\pi/3) = -\phi(\lambda)\> 
\phi(\lambda + 2\pi/3),
\end{equation}
 
\begin{equation}
\label{f19}
T^+(\lambda)\>T^+(\lambda-2\pi/3) = \phi(\lambda)\>T^-(\lambda - 2\pi/3).
\end{equation}
In order to find solutions of these last equations, 
it will be useful to  use the general relation

\begin{equation}
\label{f20}
a^L-b^L = \prod_{j=1}^L (a-\omega ^j b), \quad \omega=\exp(2\pi i/L)
\end{equation}
to write
\begin{equation}
\label{f21}
\phi (\lambda) = \sin ^L (\lambda + \pi/3) - 
\sin ^L (\lambda -\pi/3) = \prod_{l=1}^L f_l(\lambda),
\end{equation}
where
\begin{equation}
\label{f22}
f_l (\lambda) = \sin (\lambda + \pi/3) - \omega ^l \sin (\lambda -\pi/3), \quad
(l=1,..,L).
\end{equation}
 Now consider any subset $I$ of non-repeated integers of 
the set $I_0=\{1,2,...,L\}$, and the complementary 
subset $\bar {I}$ , such that $I \bigoplus \bar {I} = I_0$. 
We may try to solve (\ref{f18})-(\ref{f19}) by the ansatz
\begin{equation}
\label{f23}
T^{\pm}(\lambda)=t_0^{\pm} \prod_{l \in I} f_l 
(\lambda \pm 2\pi/3) \prod_{m \in \bar{I}} f_m (\lambda),
\end{equation}
where $t_0^{\pm}$ are unknown constants.  
This ansatz imply
\begin{eqnarray}
\label{f24}
&&T^{+}(\lambda)\>T^-(\lambda+2\pi/3)=t_0^+ t_0^- \> 
\prod_{l=1}^L f_l(\lambda +2\pi/3) \>\prod_{m=1}^L f_m(\lambda) = 
\nonumber \\
&&=t_0^+t_0^- \> \phi(\lambda + 2\pi/3 )\> \phi(\lambda),
\end{eqnarray}
where (\ref{f21}) was used in the last equality. 
The equation (\ref{f18}) imply the constraint
\begin{equation}
\label{f25}
t_0^+t_0^- =-1.
\end{equation}
Also from (\ref{f22}) and (\ref{f20}) 
\begin{eqnarray}
\label{f26}
&&T^{+}(\lambda)\>T^+(\lambda-2\pi/3)=(t_0^+)^2 
\phi (\lambda)  \prod_{l \in I} f_l(\lambda +2\pi/3) 
\prod_{m\in \bar{I}} f_m(\lambda -2\pi/3)\> = \nonumber \\
&&=(t_0^+)^2 \phi(\lambda) T^-(\lambda - 2\pi/3)/t_0^-,
\end{eqnarray}
and (\ref{f19}) imply, by using (\ref{f25}), that
\begin{equation}
\label{f27}
(t_0^-)^3=1,\quad t_0^+=-1/t_0^-.
\end{equation}
Then the ansatz (\ref{f22}) with (\ref{f25}) gives us a 
set of solutions for $T^{\pm}(\lambda)$, that when inserted 
in the matricial equations (\ref{f15}) and (\ref{f16}) will 
provide the function $Q_1(\lambda)$ and $Q_2(\lambda)$.
The zeros of these last functions are the Bethe-ansatz roots 
and the eigenenergies are calculated by using in (\ref{ba1}) 
the roots of $Q_2(\lambda)$. Instead of calculating the energies 
through this procedure, we are going to calculate them using the 
transfer matrix eigenvalues $T^-(\lambda)$.
From (\ref{ba1}) and the definition of $Q_2(\lambda)$ it is not 
difficult to obtain the relation
\begin{equation}
\label{f28}
E=\frac{\sqrt{3}}{2} \frac{d}{d\lambda} \ln \biggl(\frac{Q_2(\lambda)}
{Q_2(-\lambda)}\biggr)\biggl|_{\lambda=\pi/3} .
\end{equation}
On the other hand let us expand (\ref{f10}) with $\eta=2\pi/3$ for 
$\lambda=\eta +\epsilon \quad \epsilon \ll 1.$ 
The terms of the lowest order give us the relation
\begin{equation}
\label{f29}
 \frac{d}{d\lambda} \ln \biggl(\frac{Q_2(\lambda)}
{Q_2(-\lambda)}\biggr)\biggl|_{\lambda=\pi/3} =-\frac{L}
{\sqrt{3}}- \frac{d}{d\lambda} \ln T^-(\lambda)|_{\lambda=\pi/3},
\end{equation}
that from (\ref{f28}) provide the simple result
\begin{equation}
\label{f30}
 E=-\frac{L}{2}-\frac{\sqrt{3}}{2} \frac{d}{d\lambda} 
\ln T^-(\lambda)|_{\lambda=\pi/3},
\end{equation}
The eigenenergies associated to our solutions $T^-(\lambda)$ 
are then obtained by inserting (\ref{f23}) in (\ref{f30}), 
and we obtain after some simple algebraic manipulations
\begin{equation}
\label{f31}
 E=-L +\sum_{m \in \bar {I}} (1+2 \cos(\frac{2\pi m}{L}))=-
\sum_{l \in I} (1+2 \cos(\frac{2\pi l}{L})),
\end{equation}
where we  used the formula
\begin{equation}
 \sum_{l \in I \cup \bar {I}} (1+2 \cos(2\pi l/L))=\sum_{l=1}^L 
(1+2 \cos(2\pi l/L))=L.
\end{equation}
Also the zero-order term in the same expansion of (\ref{f10}) give us 
\begin{equation}
\label{f32}
 \frac{T^-(\pi/3)}{\sin ^L (2\pi/3)}=\frac{Q_2(\pi/3)}{Q_2(-\pi/3)}
=\prod_{k=1}^{m_2} \frac{\sin(u_k-\eta /2)}{\sin(u_k+\eta /2)}=\exp(i P),
\end{equation}
where from (\ref{ba4}) $P$ is the momentum associated to
 our solution $T^-(\lambda)$ given in (\ref{f23}). Inserting (\ref{f23}) 
into (\ref{f32}) we obtain after some simple calculations
\begin{equation}
\label{f33}
\exp(i P) =(-1)^{L+1} t^-_0 \exp\biggl(-\frac{2\pi i}{L} 
\sum_{m \in \bar {I}} m \biggr) =  t^-_0 
\exp\biggl(\frac{2\pi i}{L} \sum_{l \in I} l\biggr), \quad (t_0^-)^3=1.
\end{equation}

\vskip 1em

\begin{center}{\bf5. Analytic solutions of the Bethe ansatz equations}
\end{center}

As we discussed in the previous section, at least in the periodic case,
 the Bethe ansatz roots, corresponding to the 
eigenenergies (\ref{f31}) we observed, can be obtained   
 from the expansion of  
$Q_2(\lambda)$ given in (\ref{f1}), derived by solving (\ref{f16})
 with $T^{\pm}(\lambda)$ given by the ansatz (\ref{f23}).
Distinctly in this section we are going to present in a direct way a 
set of guessed solutions $\{u_i,v_j\}$ of the BAE that gives the 
energies conjectured in section 3. We show that they are correct by 
a direct substitution into the BAE. We present solutions of the BAE 
for the periodic and free boundaries cases. 
As we conjectured in section 3, in the case of periodic 
boundaries there exist some selection rules in the spectrum 
composition (see conjecture 1). At the end of this section we 
are going to explain partially this conjecture.

Let us consider separately the periodic and free boundary case.

\noindent{\bf 5a. Periodic case.}

\noindent  The BAE (\ref{ba2}) at $\eta=2\pi/3$, expressed in terms of the 
variables
\begin{equation}
\label{xkyl}
x_k=\frac{\sin(u_k-\pi /3)}{\sin(u_k+\pi /3)}, 
\quad y_l=\frac{\sin(v_l-\pi /3)}{\sin(v_l+\pi /3)},
\end{equation}
with $k=1,2,...,n_0+n_1$ and $l=1,2,...,n_0$ are given by
\begin{eqnarray}
\label{Bperiodic1}
&&(-1)^{n_1+1} \prod_{j^{\prime}=1}^{n_0}
\frac{1+y_j+y_j y_{j^{\prime}}}
     {1+y_{j^{\prime}}+y_j y_{j^{\prime}}}
  \prod_{k^{\prime}=1}^{n_0+n_1}
\frac{1+y_j+y_j x_{k^{\prime}}}
     {1+x_{k^{\prime}}+y_j x_{k^{\prime}}}=1, \\
&&\quad (j = 1,2, \ldots n_0) \nonumber 
\end{eqnarray}
and
\begin{eqnarray}
\label{Bperiodic2}
&&(-1)^{n_1+1} \prod_{j^{\prime}=1}^{n_0}
\frac{1+x_k+x_k y_{j^{\prime}}}
     {1+y_{j^{\prime}}+x_k y_{j^{\prime}}}
  \prod_{k^{\prime}=1}^{n_0+n_1}
\frac{1+x_k+x_k x_{k^{\prime}}}
     {1+x_{k^{\prime}}+x_k x_{k^{\prime}}}= x_k^L\\ &&
(k = 1,2, \ldots n_0+n_1). \nonumber
\end{eqnarray}
Let us fix $2 n_0+ n_1=L$. 
Our guessed solutions are obtained by considering 
$\{x_h,y_l\}$ $(k=1,\ldots,n_0+n_1,l=1,\ldots,n_0)$ 
as an arbitrary permutation of  
$\{\omega, \omega ^2,...,\omega ^L\}$, where $\omega=\exp(2\pi i/L)$.
In this case, the left side of equation  (\ref{Bperiodic1}) takes 
the form:
\begin{equation}
(-1)^{L+1} \prod_{l=1}^{L}
\frac{1+y_j+y_j \omega^l}
     {1+\omega ^l+y_j \omega^l}.
\end{equation}
Using the  identity (\ref{f20}) and the fact that $y_j^L =1$ 
we can rewrite  this product as 
\begin{equation}
(-1)^{L+1} \frac{(1+y_j)^L+(-1)^{L+1} y_j ^L}
     {1+(-1)^{L+1} (1+y_j)^L}=1\;\;.  
\end{equation}
It is evident that the 
second BAE is also satisfied due do equality
$x_k^L=1$.

Consequently we have found a subclass of solutions for the 
nested Bethe ansatz equations.
These solutions are characterized by the subset $I$ with unequal 
elements of the set 
 $I_0 =\{1,2,...,L\}$, 
and have the energy
\begin{equation}
\label{energy}
E_I=-\sum_{k=1}^{n_0+n_1} (1+x_k+x_k^{-1})=
-\sum_{l \in I} (1+2 \cos(2\pi l/L))
\end{equation}
and momentum
\begin{equation}
\label{momentum}
P_I=\sum_{k=1}^{n_0+n_1} \frac{1}{i} \ln (x_k)= 
\frac{2\pi}{L}\sum_{l \in I} l.
\end{equation}
Comparing the above relations with relations (\ref{C1}) and 
(\ref{C2}) we verify that our guessed  solutions  are 
consistent with the conjecture 1.
It is not clear however  if the corresponding wave function is not 
a  zero vector.

\noindent{\bf 5b. Free boundary case.}

\noindent  The BAE (\ref{ba3}) at $\eta=2\pi/3$, 
expressed in terms of the same variables
$x_k$ and $y_l$ 
with $k=1,2,...,n_0+n_1$ and $l=1,2,...,n_0$ are given by
\begin{eqnarray}
\label{Bfree1}
&&\prod_{j^{\prime}=1, j^{\prime} \ne j}^{n_0}
\biggl(\frac{1+y_j+y_j y_{j^{\prime}}}
     {1+y_{j^{\prime}}+y_j y_{j^{\prime}}}\biggr)
\biggl( \frac{y_j+y_{j^{\prime}}+y_j y_{j^{\prime}}}
     {1+y_j+y_{j^{\prime}}}\biggr)  \nonumber \\
&&\times \prod_{k^{\prime}=1}^{n_0+n_1}
\biggl(\frac{1+y_j+y_j x_{k^{\prime}}}
     {1+x_{k^{\prime}}+y_j x_{k^{\prime}}}\biggr)
\biggl(\frac{y_j+x_{k^{\prime}}+y_j x_{k^{\prime}}}
     {1+y_j+x_{k^{\prime}}}\biggr)=1, \\
&&\quad (j = 1,2, \ldots n_0) \nonumber 
\end{eqnarray}
and
\begin{eqnarray}
\label{Bfree2}
&&\prod_{j^{\prime}=1}^{n_0}
\biggl(\frac{1+x_k+x_k y_{j^{\prime}}}
     {1+y_{j^{\prime}}+x_k y_{j^{\prime}}}\biggr)
 \biggl(\frac{x_k+y_{j^{\prime}}+x_k y_{j^{\prime}}}
     {1+x_k + y_{j^{\prime}}}\biggr) \nonumber \\
&&\times   \prod_{k^{\prime}=1, k^{\prime} \ne k}^{n_0+n_1}
\biggl(\frac{1+x_k+x_k x_{k^{\prime}}}
     {1+x_{k^{\prime}}+x_k x_{k^{\prime}}}\biggr)
\biggl(\frac{x_k+ x_{k^{\prime}}+x_k x_{k^{\prime}}}
     {1+x_k+x_{k^{\prime}}}\biggr)= x_k^{2L}\\ 
&&(k = 1,2, \ldots n_0+n_1). \nonumber
\end{eqnarray}

Now let us  fix $2 n_0+ n_1=L-1$. 
Our guessed solutions are now given by the set 
$\{x_k,y_l\} (k=1,\ldots,n_0+n_1; l= 1, 
\ldots,n_0)$ formed by an arbitrary permutation of 
 $\{\omega, 
\omega ^2,...,\omega ^{L-1}\}$, where $\omega=\exp(i\pi /L)$.
Using the  identity
\begin{equation}
\label{forFree}
\prod_{m=1}^{L-1}\frac{(a+\omega ^m)
(1/a+\omega^m)}{(b+\omega ^m)(1/b+\omega ^m)}=
\frac{b^{L-1}}{a^{L-1}} \frac{(b^2-1)}{(a^2-1)}   
\frac{(a^{2L}-1)}{(b^{2L}-1)}, 
\end{equation}
and the fact that $y_i^L = 1$ we can easily 
verify that the BAE (\ref{Bfree1}) and 
(\ref{Bfree2}) are satisfied.

As in the periodic case,  we have found a 
subclass of solutions for the nested BAE.
These solutions are characterized 
by a subset $I \subset \{1,2,...,L-1\}$
and have the energy
\begin{equation}
\label{energyfree}
E_I=-\sum_{k=1}^{n_0+n_1} (1+x_k+x_k^{-1})=-\sum_{l \in I} 
(1+2 \cos(\pi l/L)).  
\end{equation}
All these  solutions which are consistent with conjecture  3,
so we think that corresponding Bethe wave function is not a zero 
vector.

Finally in order to conclude this section we intend to explain 
partially the selection rules formulated in 
Conjecture 1 for the periodic case. We are going to this 
by exploiting  our solutions (\ref{f23}) for 
$T^{\pm}(\lambda )$ of the functional relations of 
section 4 with the help of some ideas developed in the 
 papers~\cite{PrStr}.
  
Inserting our solutions (\ref{f23}) 
for $T^{\pm}(\lambda)$ into equation (\ref{f12}) we obtain 
\begin{eqnarray}
&&\prod_{l=1}^L f_l(\lambda)\>\> Q_1(\lambda-\pi) - t_0^+ 
\prod_{l \in I} f_l(\lambda+2\pi/3) \prod_{m \in \bar {I}} 
f_m(\lambda) \>\>Q_1(\lambda-\pi/3)+\nonumber \\
&&t_0^- \prod_{l \in I} f_l(\lambda-2\pi/3) 
\prod_{m \in \bar {I}} f_m(\lambda) \>\>Q_1(\lambda+\pi/3)=0.
\end{eqnarray}
Dividing by the common factor $\prod_{m \in \bar {I}} 
f_m(\lambda)$ we obtain 
\begin{equation}
\label{f34}
F_1(\lambda)\>\> Q_1(\lambda-\pi) + \Omega \>
F_1(\lambda+2\pi/3)\>\>Q_1(\lambda-\pi/3)+ 
\Omega^2 F_1(\lambda-2\pi/3)\>\>Q_1(\lambda+\pi/3) = 0,
\end{equation}
where $\Omega=-t_0^+$ ($\Omega^3=1$) and
\begin{equation}
\label{fF}
F_1(\lambda)=\prod_{l \in I} f_l(\lambda).  
\end{equation}
On the other hand the solution (\ref{f23}) for 
$T^{\pm}(\lambda )$ brings (\ref{f12}) into 
a similar functional equation: 

\begin{equation}
\label{f35}
F_2(\lambda)\>\> Q_2(\lambda-\pi) + 
\Omega^2 \>F_2(\lambda+2\pi/3)\>\>Q_2(\lambda-\pi/3)+ 
\Omega F_2(\lambda-2\pi/3)\>\>Q_2(\lambda+\pi/3)= 0,
\end{equation}
where 
\begin{equation}
\label{fPhi}
F_2(\lambda)= \sin ^L \lambda \prod_{m \in \bar {I}} f_m(\lambda).  
\end{equation}

Let us consider the case where $L \neq 3n$. In this case 
 since $P = \frac{2\pi}{L}j$ ($j =0,\ldots,L-1$), 
equation (\ref{f33}) gives 
$t_0^- =1$, and consequently $\Omega =1$ in 
(\ref{f34}) and (\ref{f35}).

We intend to argue now that there exist pairs $\{Q_1(\lambda), Q_2(\lambda)\}$,
satisfying  (\ref{f34}) and (\ref{f35}) 
with $\Omega=1$ which lead to "physical" solutions for 
the nested BAE (\ref{ba2}), i. e.,  solutions which are 
inside the usual bounds $n_0 \leq n_1 \leq n_2$ or equivalently 
  
\begin{equation}
\label{bounds}
\mbox{deg}\>Q_1 \leq  \mbox{deg}\>Q_2 - \mbox{deg}\>Q_1 
\leq L - \mbox{deg}\>Q_2.
\end{equation}
First of all, we have special solutions for  (\ref{f34}) and (\ref{f35})
which can be written as

\begin{equation}
\label{specsol}
Q_1(\lambda) = Q_{1spec}(\lambda) = 
\prod_{m \in \bar{I}} f_m(\lambda + \pi), \quad
Q_2(\lambda) = Q_{2spec}(\lambda)= 
\prod_{l \in I} f_l(\lambda + \pi)
\end{equation}

Let us check these formulas inserting them into equations (\ref{f34}) and (\ref{f35}). The left side of equation (\ref{f34}) becomes (see (\ref{f21}))

\begin{eqnarray}
&&\prod_{l \in I} f_l(\lambda)\> \prod_{m \in \bar{I}} f_m(\lambda)\>+
\prod_{l \in I} f_l(\lambda+2\pi/3)\> \prod_{m \in \bar{I}} f_m(\lambda+2\pi/3)\>+
\nonumber \\
&&
+\prod_{l \in I} f_l(\lambda-2\pi/3)\> \prod_{m \in \bar{I}} f_m(\lambda-2\pi/3)\>= \nonumber \\
&&=\sin^L(\lambda + \pi/3) - \sin^L(\lambda - \pi/3) + \sin^L(\lambda + \pi) - \nonumber \\ 
&&- \sin^L(\lambda + \pi/3) + \sin^L(\lambda - \pi/3) -
\sin^L(\lambda - \pi) =0. \nonumber
\end{eqnarray}
Similarly the left side of equation (\ref{f35}) becomes
\begin{eqnarray}
&&\sin^L(\lambda)\>\>\>\> (\sin^L(\lambda + \pi/3) - \sin^L(\lambda - \pi/3)) + \nonumber \\
&&+\sin^L(\lambda+2\pi/3)\>\> (\sin^L(\lambda + \pi) - \sin^L(\lambda + \pi/3)) + \nonumber \\ 
&&+\sin^L(\lambda-2\pi/3)\>\> ( \sin^L(\lambda - \pi/3) -
\sin^L(\lambda - \pi)) =0. \nonumber
\end{eqnarray}

Let $0 \le {\cal I} \le L$ is the number of elements of the set $I$. 
Then degrees of these special solutions $Q_1$ and $Q_2$ are $L-{\cal I}$ and
${\cal I}$ respectively.
Inequalities (\ref{bounds}) for these pairs become 

\begin{equation}
L-{\cal I} \leq 2 {\cal I} - L \leq L- {\cal I},
\end{equation}
which is the same as the equality $2L=3{\cal I}$. It is not enough for our purposes, especially for $L \ne 3n$, so we have to look for additional solutions. They do exist due to the fact that the matrices in equations (\ref{f15}) and (\ref{f16}) for $Q_1$ and $Q_2$ has rank 1.

According to the analysis of functional equations of type 
(\ref{f34}) or (\ref{f35}) made in some previous papers~\cite{PrStr} 
it was noticed that equations of this type have some conjectured 
properties that we are going to accept. If in (\ref{f34}) or 
(\ref{f35}) $F_i(\lambda)$ ($i=1,2$) have a trigonometric 
form 
$F_i(\lambda) = \prod_{j=1}^N \sin (\lambda - a_j)$, 
of degree $N$, in general there exists a trigonometric 
solution of the form 
$Q_i(\lambda) = \prod_{j=1}^m \sin (\lambda - b_j)$ of degree 
$m$. This degree depends on the value of $\Omega$ 
appearing in the equation. In particular if 
$\Omega=1$ then $m = N/2 +1$ for $N$ even and $m = (N-1)/2$ for 
$N$ odd. Only  for special choices of 
$F_i(\lambda)$ this degree can 
be decreased. 
We call these solutions $Q_{1gen},Q_{2gen}$ as general ones.

 Due to 
(\ref{fF}) and (\ref{fPhi}) we 
have $\mbox{deg}\>F_1(\lambda)= \cal I $ 
and $\mbox{deg}\>F_2 (\lambda) =2L-\cal I $. 
If we chose  $\cal I $ even  then $2L-\cal I$ 
is also even and  the equations (\ref{f34}) 
and (\ref{f35}) have trigonometric solutions 
for $Q_1$ and $Q_2$, with 
$\mbox{deg} Q_1 = {\cal I}/2 +1$ and $\mbox{deg} 
Q_2 =(2L - {\cal I})/2 +1$. 
On the other hand for odd values of $\cal I$ we have  
$\mbox{deg} Q_1 = ({\cal I} -1)/2$ and $\mbox{deg} 
Q_2 = (2L -{\cal I} -1)/2$. 

Before consider arbitrary values of $L$ let us restrict  initially 
 to the particular case $L=7$. 
 In table 1 we list the predicted degrees of the sine-polynomials 
$Q_1$ and $Q_2$. 
We underline pairs $Q_1, Q_2$ which satisfy the 
inequalities (\ref{bounds}) and in the last column
of this table we present the eigensectors  
where we expect to find the  predicted simple energy levels.

\vspace{0.5cm}

\begin{tabular}{|c|c|c|c|c|c|}
\hline
${\cal I}$ & deg $Q_{1gen}$  & deg $Q_{2spec}$ & deg $Q_{1spec}$ & deg $Q_{2gen}$ & sector \\
\hline
0 & 1 & 0 & 7 & 8 & - \\
1 & \underline{0} & \underline{1} & 6 & 6 & (0,1,6) \\
2 & 2 & 2 & 5 & 7 & - \\
3 & \underline{1} & \underline{3} & 4 & 5 & (1,2,4) \\
4 & 3 & 4 & 3 & 6 & - \\
5 & 2 & 5 & \underline{2} & \underline{4} & (2,2,3) \\
6 & 4 & 6 & 1 & 5 & - \\
7 & 3 & 7 & \underline{0} & \underline{3} & (0,3,4) \\
\hline
\end{tabular}

\vspace{0.5cm}

First of all we see that only odd ${\cal I}$ leads to "physical" solution.
This fact is consistent with the results of our 
"experimental" observations formulated in Conjecture 1.

We see further that for small ${\cal I}$ the "physical" solution is a pair consisting of a general solution $Q_{1gen}$ and a special one $Q_{2spec}$. 
For odd ${\cal I}=2k+1$ we have deg $Q_{1gen}=({\cal I}-1)/2=k$
and deg $Q_{2spec}={\cal I}=2k+1$. Inserting these formulas into inequalities 
(\ref{bounds}) we get $k \leq k+1 \leq L-2k-1$. For $L=3n+l\>\> (l=1,2)$ we obtain the upper boundary for k:

\begin{equation}
\label{upperb}
k \leq n+\frac{l-2}{3}
\end{equation}

On  the other side for ${\cal I}$ large enough we combine a special solution
$Q_{1spec}$, which has degree $L-{\cal I}=L-2k-1$ and a general one $Q_{2gen}$ which has degree $(2L-{\cal I}-1)/2=L-k-1$.
Inequalities (\ref{bounds}) become $L-2k-1 \leq k \leq k+1$ 
 Taking  $L=3n+l\>\> (l=1,2)$ we obtain now the lower boundary for k:

\begin{equation}
\label{lowerb}
k \geq n+\frac{l-1}{3}
\end{equation}

There is no  holes between (\ref{upperb}) and 
(\ref{lowerb}) so we have "physical" solution for every odd ${\cal I}$ and corresponding energy level have to be in sector $(k,k+1,L-2k-1)$.  This explains Conjecture 1!

The case  $L=3n$ is more complicated 
and we did not derive similar results. 

\vskip 1cm

\begin{center}{\bf6. Summary and Conclusions}
\end{center}

Although the exact integrability is a property independent 
of the lattice size, the exact solution of the associated 
BAE for finite chains were know in very few cases. 
The XXZ chain at the special value of the anisotropy 
$\Delta = (q + q^{-1})/2$, $q = \exp(i2\pi/3)$ 
is one of these examples. Motivated by this result we 
made extensive numerical calculations 
for the $SU(3)$ generalization of the XXZ chain, namely the 
$SU(3)$ Perk-Schultz model, also at the special 
anisotropy $q= \exp(i2\pi/3)$. 
Surprisingly, as we stated in section 3, the numerical 
results reveal that many of the eigenenergies (not all of 
them) are expressed as combinations of simple 
cosines and, apart from some selection rules, are quite 
similar to the energies of a free fermion chain (or 
XXZ at $\Delta = 0$).

Our numerical results indicate the five conjectures presented 
in section 3. The first two conjectures concerns with 
the periodic quantum chain and gives the exact expression for the 
energy and momentum of several eigenfunctions. In several 
sectors the lowest energy value is also predicted. 
In order to explain these results analytically we present 
in section 5 a set of BAE solutions that are consistent 
with the conjectured energies.  However the set of 
solutions we obtained is larger then the conjectured one. 
This imply that some of our solutions, although having 
non-coinciding  roots are unphysical, corresponding 
to a zero vector, since the associated energy is 
missing in the eigenspectrum. 
These missing BAE solutions appear in the sectors  
($n_0, n_1,n_2$) not satisfying the 
bound $n_0 \leq n_1 \leq n_2 \leq 2L/3$. From the functional 
relations derived in section 4 we were able to explain 
at least for the cases $L \neq 3n$ the selection rules 
appearing in the conjecture 1. In the case $L=3n$ the degree 
of the trigonometric solutions of the functional 
equations are more difficult to predict and we could not explain 
the conjecture 1.

	The last three conjectures concerns the eigenspectra of the 
Hamiltonian with the quantum symmetry $SU(3)_q$, i. e., free 
boundary case. These conjectures shows no 
selection rules in contrast with the periodic case. 
Again in this case we present a set of solutions of the 
BAE sharing the same energies as those of conjecture 3. 
The functional relations in this case are more complicated and 
we leave this analysis for a  future work.

Finally it is interesting to mention that the finite-size 
corrections obtained from the conjectured eigenenergies of section 3 
give us some conformal dimensions of the underlying conformal 
field theory (CFT) governing the large-distance physics of the 
model. As a consequence of the conformal invariance of 
the infinite system these eigenenergies \cite{cardy} should 
behave as 
\begin{equation} \label{A}
E = e_{\infty}L + \frac{\pi}{6L} v_s (12x_o - c) + o(L^{-1}),
\end{equation}
in the periodic case, and 
\begin{equation} \label{B}
E = e_{\infty}L +f_s  + \frac{\pi}{24L} v_s (24x_o^s - c)+ o(L^{-1}),
\end{equation}
in the open boundary cases. In the above expression 
$e_{\infty}$ and $f_s$ are the energy per site and surface 
energy in the bulk limit, $v_s$ is the sound velocity, $c$ is 
the central charge and $x_o, x_o^s$ are the conformal 
dimensions governing the power-law decay of 
correlations in the periodic and open chain cases.

In the periodic case, the conjecture 2 gives the asymptotic 
behaviour for the lowest eigenenergies
\begin{equation} \label{C}
E = e_{\infty} L -\frac{\pi}{6L}v_s(-2)+ O(L^{-3})  \quad \mbox{for} \quad 
L=3n,
\end{equation}
\begin{equation} \label{D}
E = e_{\infty} L -\frac{\pi}{6L}v_s\frac{2}{3}+O(L^{-3})  \quad \mbox{for} 
\quad L \neq 3n,  
\end{equation}
where $e_{\infty} = -(\frac{2}{3} + \frac{\sqrt{3}}{\pi})$ and 
$v_s = \sqrt{3}$ can be inferred from the lowest eigenenergy 
with momentum $\frac{2\pi}{L}$ 
of conjecture 1. 

The underlying $U(1)\otimes U(1)$ CFT governing these quantum 
chains is expected to have a central charge 
$c=2$ and when formulated in the torus should have 
the conformal dimensions \cite{dim}
\begin{equation} \label{E}
x(n_1,n_2;m_1,m_2) = x_p(n_1^2-n_1n_2+n_2^2) + 
\frac{1}{12x_p}(m_1^2+m_1m_2 +m_2^2),
\end{equation}
where $x_p$ is related 
 with the compactification 
ratio and $n_1,n_2,m_1,m_2$ are expected to be integers.
Assuming $c=2$ in (\ref{A}) and comparing with 
relations (\ref{C}) and (\ref{D}) we obtain for 
the predicted lowest eigenenergies,  the 
associated dimensions $x=\frac{1}{3}$ for $L=3n$ and 
$x=\frac{1}{9}$ for $L \neq 3n$. 
From (\ref{E}) these dimensions can be identified with 
$x(1,1;0,0) = x_p = \frac{1}{3}$ and 
$x(\frac{1}{3},-\frac{1}{3};0,0) = \frac{x_p}{3} =\frac{ 1}{9}$, by 
taking $x_p = \frac{\eta}{2\pi} = 
\frac{1}{3}$.   The fractional values in the 
last case happens because the ground state for lattices 
with sizes $L\neq 3n$
does not represent, in the bulk limit, the true vacuum 
of the CFT, since it contains topological defects.

In the case of open boundaries conjecture 5 give us 
for the ground state 
\begin{equation} \label{F}
E = e_{\infty}L +f_s - \frac{\pi}{24L} v_s (-2) +O(L^{-3}),
\end{equation}
where $e_{\infty}$ and $v_s$ was already obtained in the 
periodic case and $f_s =3/2$. Comparing (\ref{F})
with (\ref{B}) we obtain $c=-2$. This can be understood 
since the quantum chain with open boundaries is 
$SU(3)_q$ symmetric, with $q=e^{i\eta}$, $\eta=\frac{2\pi}{3}$ 
and the 
expected \cite{reduct} conformal anomaly  in this case is 
$c=2 -\frac{ 24}{m(m+1)}$, where $m =\frac{ \eta}{\pi-\eta} = 2$. 
Similar analysis can also be done for the excited states.

{\it Acknowledgments} We thank  A. V. 
Razumov for useful discussions. This work was supported 
in part by the brazilian agencies FAPESP and CNPQ (Brazil), and by the Grant 
 \# 01--01--00201 (Russia).

\newpage

\begin{center} 
  	TABLE CAPTION
\end{center}
Table 1 - Degrees of the polynomials $Q_1$ and $Q_2$ coming 
from the possible solutions for $L+7$. The special solutions 
$Q_{1spec}$ and $Q_{2spec}$ are given by (\ref{specsol})
and the general ones $Q_{1gen}$ and $Q_{2gen}$ are discussed 
in the text. 

\end{document}